\documentclass[12pt, preprint]{aastex}
\shorttitle{DIFFUSE UV EMISSION FROM DRACO}
\shortauthors{Sujatha et al.}

\newcommand{\phunit}{photons cm$^{-2}$ sr$^{-1}$ s$^{-1}$ \AA$^{-1}$}

\newcommand{\iras} {{\it IRAS}}
\newcommand{\galex} {{\it GALEX}}

\begin{document}
\title{ \galex\ OBSERVATIONS OF DIFFUSE ULTRAVIOLET EMISSION FROM DRACO}
\author{N. V. SUJATHA, JAYANT MURTHY}
\affil{Indian Institute of Astrophysics, Koramangala, Bangalore - 560 034, India}
\email{sujatha@iiap.res.in, murthy@iiap.res.in}

\author{RAHUL SURESH}
\affil{National Institute of Technology Karnataka, Surathkal, Managalore - 575 025, India}
\email{rsbullz@gmail.com}

\author{RICHARD CONN HENRY \and LUCIANA BIANCHI}
\affil{Department of Physics and Astronomy, The Johns Hopkins University,\\
 Baltimore, MD 21218, USA}
\email{rch@pha.jhu.edu, bianchi@pha.jhu.edu}

\begin{abstract}
We have studied small scale (2\arcmin) spatial variation of the diffuse UV radiation using a set of 11 \galex\ deep observations in the constellation of Draco. We find a good correlation between the observed UV background and the IR 100~\micron\ flux, indicating that the dominant contributor of the diffuse background in the field is the scattered starlight from the interstellar dust grains. We also find strong evidence of additional emission in the FUV band which is absent in the NUV band. This is most likely due to Lyman band emission from molecular hydrogen in a ridge of dust running through the field and to line emissions from species such as C~{\small IV} (1550~\AA) and Si~{\small II} (1533~\AA) in the rest of the field. A strong correlation exists between the FUV/NUV ratio and the FUV intensity in the excess emission regions in the FUV band irrespective of the optical depth of the region. The optical depth increases more rapidly in the UV than the IR and we find that the UV/IR ratio drops off exponentially with increasing IR due to saturation effects in the UV. Using the positional details of {\it Spitzer} extragalactic objects, we find that the contribution of extragalactic light in the diffuse NUV background is 49 $\pm$ 13 \phunit\ and is 30 $\pm$ 10 \phunit\ in the FUV band.
\end{abstract}


\section{INTRODUCTION}

Studies of the diffuse ultraviolet (UV) sky have been an important part of interstellar dust studies over the last four decades \citep{Bowyer91,RCH91,JM09} but were limited by the difficulty of observing faint diffuse sources near the limit of the instrumental sensitivity. It has been generally agreed that the low and mid-latitude diffuse radiation is dominated by the scattering of starlight by interstellar dust but with a baseline at high galactic latitudes, which was variously attributed to either high latitude dust \citep{Bowyer91} or to an extragalactic source \citep{RCH02}.

Just as the {\it Infrared Astronomy Satellite} (\iras) revolutionized the study of the diffuse infrared (IR) emission \citep{Low}, data from the {\it Galaxy Evolution Explorer} (\galex) have the 
potential to change our view of the diffuse UV sky. We have begun an ambitious
effort to map the diffuse background in all \galex\ deep observations (exposure time~$\geq$~5000~sec) with the first of these being observations of a region of nebulosity first observed by \citet{Sandage76} (hereafter `Region I'), later identified as a nearby molecular cloud MBM~30 \citep{MBM}. This region has a comparatively high optical depth in the UV (0.8~$\le\ \tau \le$ 3.3) and we found a flat UV emission \citep{SNV09} despite the IR 100 \micron\ emission increasing by a factor of 2.

In this work, we examine a set of observations of a region in Draco, where the optical depth is much lower ($\tau$ $<$ 0.5) but where there is a ridge of dust extending through the field. As with the Region I observations of \citet{SNV09}, this field is at high Galactic latitude (33\degr\ - 37\degr) but is about 60 degrees away at a longitude of about 88\degr. The data are from the \galex\ Deep Imaging Survey (DIS), a few of them overlapping with the {\it Spitzer} First Look survey\footnote{See http://ssc.spitzer.caltech.edu/fls/}. Combining these two studies we present here the nature of diffuse UV radiation from low optical depth to high optical depth region.

\section{OBSERVATION AND DATA ANALYSIS}

The \galex\ spacecraft was launched in 2003 under NASA's Small Explorer (SMEX) program 
with a primary science objective of observing star formation in galaxies at low redshifts \citep{Martin2005}. Light from the sky is collected through a single 50 cm telescope and  separated into two bands (far ultraviolet (FUV): 1350 - 1750 \AA;  near ultraviolet (NUV): 1750 - 2850~\AA) using a dichroic mirror. Independent low noise delay-line detectors record every photon in each band with an overall effective spatial resolution of 5 -- 7\arcsec\ in the sky over a 1.25\degr\ field. The data products from the mission include, amongst other files, Flexible Image Transport System (FITS) \citep{Fits} images of the FUV and NUV fields and a list of point sources in each field. A complete description of the data processing, the calibration and the data products may be found in \citet{Morrissey2007}.

This work follows our study \citep{SNV09} on {\it GALEX} observations of diffuse emission in Region I and focuses on a set of 11 observations covering an area about 10 square degrees in the constellation of Draco, with cumulative exposure times of 3,000 to 50,000 seconds (Table~{\ref{obs_log}). These observations were taken by the \galex\ team as part of a program to map the {\it Space Infrared Telescope Facility} ({\it SIRTF}: now the {\it Spitzer Space Telescope}) First Light locations - hence the target name of ``{\it SIRTFFL}''. This region (Fig.~\ref{IR_img}) contains the high velocity cloud (HVC) Complex C \citep{Miville05} at a distance of more than 800 pc but also, more relevant to our data, the nearby (60 pc) cloud LVC 88+36-2 \citep{Lilienthal}, seen as a ridge in the IR emission. This cloud was first discovered to cast a shadow in the X-ray background \citep{Burrows91}. Because of the then upcoming Spitzer observations, \citet{Lockman05} mapped the region in the 21 cm line of H~{\small I}, finding several components (Table {\ref{h1_comp}). This wealth of detail has proven invaluable to our understanding of the UV observations.

Each observation is comprised of a number of visits spread over a period of months, or even 
years, all of which are coadded by the standard \galex\ pipeline \citep{Morrissey2007} to produce a single image in each of the two bands. Point sources in each image were extracted by the \galex\ team using a standard point source extractor (SExtractor - \citet{SEXT}) and a 
merged point source catalog was created. We note here that the exposure time in the FUV 
detector was often significantly less than that in the NUV because of intermittent power supply problems. Our processing uses the FITS image files and the merged point source catalog from the \galex\ pipeline. These image files have been fully calibrated and flat fielded but not background subtracted. Although the \galex\ program does provide files containing the background in each observation, these were made by fitting a multi-dimensional surface to the image and therefore show structure related to the pinning points of the surface. While perhaps adequate for their intended purpose of subtracting the background from point sources in the field, they introduce large scale artifacts which make them unsuitable for the study of the diffuse radiation field.

Following \citet{SNV09}, we created our own background files for each observation by blanking out the point sources in the merged \galex\ point source catalog and binning the observation into 2\arcmin\ pixels (80 $\times$ 80 \galex\ pixels). These images form the starting point of our analysis. Because of edge effects, we only used the central 1.15\degr\ of the 1.25\degr\ field of view for the analysis, rejecting about 20\% of the total number of pixels. These background files are comprised of the foreground emission (instrumental dark count, airglow and zodiacal light) and the astrophysical signal (atomic and molecular emission, dust scattered starlight and any extragalactic contribution).

\section{FOREGROUND EMISSION}
A large field of view imager such as \galex\ has distinct advantages in observations of the diffuse background in that stars can be easily identified and rejected. However, without spectra, we can only infer the contribution of the different components of the diffuse radiation field. Instrumental dark count is negligible, contributing less than 5 \phunit\ in either band \citep{Morrissey2007} but airglow, primarily due to the O~{\small I} lines at 1356~\AA\ and 2471~\AA, is expected to contribute about 200 \phunit\ to either band \citep{Boffi2007}. Although we cannot extract the airglow contribution directly, we have been able to use the Telemetered Event Counter (TEC) of the spacecraft to track the total number of counts as a function of orbital time. Assuming the time dependent part of TEC present in both the \galex\ bands as the total foreground emissions, a baseline has been subtracted from each visit so that the count rate is zero at local midnight. The remaining variable component of airglow (AG$_{v}$) is well fit with a quadratic as a function of time from local midnight (Fig.~\ref{TEC_time}).

In addition, we have found that the baseline levels at local midnight are strongly correlated with the 10.7 cm solar flux\footnote{http://www.dxlc.com, http://www.spaceweather.ca} (Fig.~\ref{solar_flux}) which is used as a proxy for solar-terrestrial interactions \citep{Chatterjee}. Each observation is comprised of several visits, each of which may have a different airglow and zodiacal light contribution. We have estimated and subtracted the zodiacal light from each visit's baseline level and found the y-intercept for each observation, corresponding to stars in the field and the diffuse cosmic background. These values have been subtracted from the individual baseline levels and the resultant values, assumed as the constant airglow (AG$_{c}$) in each visit, are plotted in Fig.~\ref{solar_flux}. Combining these two results (i.e., AG$_{c}$ + AG$_{v}$) allow us to calculate the total airglow (AG) as a function of local time (t, hours from local midnight) and solar 10.7 cm flux (SF, in 10$^{4}$~Jy) with the following equations, with an uncertainty of about 50 \phunit.
\begin{eqnarray}
FUV AG = 3.4\ SF + 24.5\ t^2 + 11.6\ t \\
NUV AG = 3.7\ SF + 16.1\ t^2 + 5.9\ t
\end{eqnarray}
This emission is consistent with an origin of the airglow in solar photons resonantly scattered from geocoronal oxygen atoms (L. J. Paxton, personal communication). It should, however, be noted that \citet{Brune} observed a much lower level of airglow emission with a scaled \galex\ contribution of about 50 \phunit\ from their rocket-borne spectroscopic observation. It is possible that some part of what we have euphemistically called ``airglow'', may be due to some other contributor \citep{RCH2010}.

The remaining foreground contributor, zodiacal light, is important only in the NUV band because of the rapidly fading solar spectrum at wavelengths shorter than 2000~\AA. Although 
there is no UV map of the zodiacal light, we have used the distribution in the visible with grey scattering \citep{Leinert98} to predict the zodiacal light in each visit\footnote {calculator at http://tauvex.iiap.res.in/htmls/tools/zodicalc/}. The foreground emission (Table~\ref{ag_zl}), ranges from 20\% to 50\% of the total emission with an uncertainty of about 30 \phunit, estimated using the spatial overlap between different observations. It should be emphasized that the foreground emission affects only the level of the offset and will not affect the spatial variability of the diffuse radiation field.

\subsection{Scatter in the Data}

More interesting is the scatter in the data. For a photon counting instrument such as \galex, 
the instrumental scatter will be either due to photon noise or to errors in the flat fielding (calibration) of the instrument. We have empirically derived the instrumental scatter by dividing each observation into two sets of visits, which may well be separated by several months. There is excellent agreement between this and the intrinsic photon noise (Fig. \ref{scat_plot}), confirming that the errors are dominated by poissonian rather than instrumental effects. As an independent test, we also took the overlap regions between different observations and calculated the scatter between them. Although the scatter for the overlap regions is somewhat higher than the calculated values, this is due to the many fewer points in the overlap regions and their location near the edge of the detector. We note here that all our comparisons are in sky coordinates because there are arbitrary roll angle differences between different visits, which do not allow a comparison between physical detector pixels.

\section{RESULTS AND DISCUSSION}

The FUV and NUV images of the {\it Spitzer} ``First Look'' field obtained after subtraction of the foreground emission are shown in Fig.~\ref{diffuse_image} at a spatial resolution of 2\arcmin. The UV images of Fig.~\ref{diffuse_image} may be compared with the IR 100 \micron\ map (Fig.~\ref{IR_img}). There are several possible contributors to the astrophysical UV emission, a significant one being, dust-scattered starlight which contributes to both the FUV and the NUV bands. This is reflected in the good correlation between the FUV and NUV bands (Fig.~\ref{fuv_nuv}) and between the two UV bands and the IR 100 \micron\ fluxes (Fig.~\ref{UV_IR}). This is in contrast with the essentially flat UV-IR curves obtained by \citet{SNV09} in Region I. The IR emission is due to thermal radiation from an optically thin layer of dust, as the cross-section of the grains is low in the IR. On the other hand, the cross-section of the grains is much higher in the UV and the optical depth transitions from being optically thin in these Draco observations to being optically thick in Region I. 

In Fig.~\ref{UV_IR_ratio}, we have plotted the ratio between the UV bands and the IR to understand the nature of diffuse UV emission with optical depth. There is a clear trend visible from the low optical depth Draco region to the high optical depth (in the UV) Region I with an empirical formula of $$\frac{F_{UV}}{F_{IR}} = 415\ e^{-0.22 \times F_{IR}}.$$ It is interesting to note that the $F_{UV}/F_{IR}$ ratio in our \galex\ data follows a continuous curve very similar to that found by \citet{JM01} in Orion using data from the Midcourse Space Experiment (MSX) even though the UV and the IR fluxes in Orion were each greater by a factor of about 200, reflecting the intense radiation field there. However, quite different values are cited in the literature for other regions with ratios ranging from near -50 to almost 260 \phunit(MJy sr$^{-1}$)$^{-1}$ with little dependence on the IR (\citet{Sasseen95, Sasseen96}). It is likely that these relations are only apparent when observed at a high enough spatial resolution; the MSX data were at a resolution of 20\arcsec\ and our data are at a resolution of 2\arcmin, while the other observations are at resolutions of 0.5\degr\ or worse. Since, both the IR and the UV vary on smaller scales, the measured $F_{UV}/F_{IR}$ ratio may not be a reliable estimator of the true ratio. In fact, \citet{Sasseen96} found a $F_{UV}/F_{IR}$ ratio of 255 \phunit(MJy sr$^{-1}$)$^{-1}$ for the slope using all their data, higher than any of the individual data sets. In general, we conclude that the $F_{UV}/F_{IR}$ ratio in any region  strongly depends on the local effects such as the proximity of hot stars near the scattering dust and the optical depth.

Readily apparent in both Fig.~\ref{UV_IR} and Fig.~\ref{UV_IR_ratio} is the ridge of dust (LVC 88+36-2) running through our field, where the FUV emission is proportionately greater than the NUV. Indeed, this reflects a general increase in the FUV/NUV ratio with the FUV surface brightness (Fig.~\ref{ratio_fuv}) seen here and in Region I. The most likely explanation for this is that there is an additional component in the FUV band which is not seen in the NUV. \citet{SNV09} suggested that this is fluorescent Lyman band (1400 - 1700 \AA) emission of molecular hydrogen, a reasonable assumption in Region I where \citet{MHB90} had already observed widespread H$_{2}$ fluorescent emission.

Assuming that the FUV/NUV ratio for dust scattering alone is constant with a value of 0.8 (Fig.~\ref{ratio_fuv}), we can estimate the level of excess emission in the field. The average error in this ratio, due to the scatter in the data, is estimated to be $\pm$0.12. Although the excess emission level in the field is not generally correlated with N(H~{\small I}) (Fig.~\ref{lvc-h2f}), there is a strong correlation in the ridge (LVC 88+36-2), where the excess emission is likely due to H$_{2}$ fluorescence. We obtain a reasonable fit to the data following \citet{MHB90} and calculate the emission assuming a plane-parallel slab with constant density (Fig.~\ref{H2_ridge}). \citet{Park2009} have observed atomic emission lines of both Si~{\small II} (1533 \AA) and C~{\small IV} (1550 \AA) around the nearby Draco molecular cloud which would effectively contribute about 50 \phunit\ in the FUV band and it may be that some part of the emission outside the ridge, where there is no correlation with H~{\small I}, may be due to atomic lines instead.

\subsection{Modeling the Dust Scattered Emission}

We have applied our standard three-parameter model of interstellar dust scattering \citep{SNV05} to the continuum dust scattered light in Draco. This model has been described fully by \citet{SNV05} and uses Kurucz models \citep{Kurucz} for the stars in the Hipparcos catalog \citep{Hipparcos} to calculate the interstellar radiation field \citep{SNV04}. This radiation is then scattered from dust in the line of sight, taking into account self-extinction. The scattering function is from  \citet{HG} and depends only on the albedo ($a$) and the phase function asymmetry factor ($g =\ <cos(\theta)>$). Typical values for these suggest moderately reflective ($a = 0.4$), highly forward scattering ($g = 0.6$) grains in the UV, in agreement with the predictions for a mixture of spherical carbonaceous and silicate grains \citep{Dr03}. On account of the uncertainity of extragalactic contribution (EGL) in the data, we have considered it as a variable parameter in the model. A full treatment of the problem would take into account multiple scattering and clumpiness in the ISM \citep[see, for example,][]{Gord04} but, because the optical depth is low ($\tau$ $<$ 0.5) in our observations, we have used a single scattering model with no clumping. Correlation studies between the UV emissions and different components of H~{\small I} in the region show that the diffuse emission is correlated maximum with the LVC component of H~{\small I}, which is the local cloud at 60 pc, and the addition of any other components such as IVC or HVC to LVC reduces the correlation. The details (correlation coefficient, $r$) are given in Table~\ref{correl}. Hence for these observations, we have assumed scattering from the local clouds at a distance of 60 pc; very little contribution to the diffuse light comes from the more distant clouds.

With these assumptions, we have placed 1$\sigma$ limits of 0.45~$\pm$~0.08 on the albedo ($a$), 0.56~$\pm$~0.10 on $g$ and 58~$\pm$~18 \phunit\ on the EGL in the NUV band with a reduced $\chi^{2}$ of 1.32. If we use the empirical ratio of 0.8 for the FUV/NUV ratio of the dust, the best fit NUV values translate into an albedo of 0.32~$\pm$~0.09 and $g$ of 0.51~$\pm$~0.19 in the FUV. These results are in reasonable agreement with previously determined values \citep{Dr03}. The scatter in our data is more than can be accounted for by photon noise alone.  We have empirically derived a 1$\sigma$ error bar of about 40~\phunit\ in the model fit to the data compared to about 20~\phunit\ from the photon noise,  probably reflecting the incompleteness of the model.

\subsection{Contribution from Extragalactic Objects}

{\it Spitzer Space Telescope} \citep{Werner} made its 67 hours First Look Survey (FLS) near 
Draco in 2003 in order to characterize the starlight from distant galaxies in the region in mid-infrared, using {\it Infrared Array Camera} ({\it IRAC};~\citet{Fazio}) and the {\it Multiband Imaging Photometer for Spitzer} ({\it MIPS};~\citet{Rieke}). The {\it IRAC} survey covered an area of 3.8 deg$^{2}$ centered on R.A. 17$^{h}$18$^{m}$00$^{s}$, Dec. +59\degr30\arcmin00\arcsec\ at wavelengths 3.6, 4.5, 5.8 and 8.0 micron, with flux density limits of 20, 25, 100 and 100 $\mu$Jy \citep{Lacy}. This instrument produced a band merged catalog of the survey containing 103,193 objects with a positional accuracy of about 0.25\arcsec\ for high signal-to-noise objects and about 1\arcsec\ at the flux density limits. The overlap area of {\it IRAC} survey is about 38\% of the total \galex\ observed area in Draco. We have used this important positional details of {\it IRAC}  cataloged sources to estimate the observed EGL contribution in our diffuse maps. Note that the only expected contribution of EGL in our diffuse maps are from the undetected faint galaxies by SExtractor, since we have removed all the detected sources using the \galex\ catalog from each of our field. 

We find that some {\it IRAC} objects are showing enhancement in the UV intensities from their local background, measured from 2\arcmin\ bin. In Fig.~\ref{uv_enh}, the average UV intensities of these objects measured using a diameter of 9\arcsec\ (6 pixels) are plotted against the corresponding local background. The UV intensities and the corresponding AB magnitudes of these sources are estimated after subtracting the local background. The total number of such objects detected in the NUV field is 18,989 in the magnitude range 20.0 -- 24.0. The number counts of these objects (Table~\ref{num_cts_tab}) are shown in Fig.\ref{num_cts}. By integrating along the curve, we derived the EGL contribution in the NUV map as 49~$\pm$~13 \phunit. The errorbar include both the uncertainities in the magnitude and the area overlapped by {\it IRAC} in the field.  It is interesting to note that this amount is in good agreement with the extracted value of EGL from the model. We have also found that an accurate estimation of number counts in the FUV band is difficult due to the excess emission in the field and hence we restricted our analysis to the NUV band. However, assuming an average ratio of 0.43 between the FUV and NUV sources derived from \citet{Xu2005} and \citet{Hammer2010} in the magnitude range 20.0 -- 24.0,  we estimated the EGL contribution as 30~$\pm$~10 \phunit\ in the FUV map from a total of 8165 objects.

\section{CONCLUSIONS}

We have completed an analysis of two sets of deep \galex\ observations: earlier near the Sandage reflection nebulosity (Region I) towards MBM 30 \citep{SNV09} and now near the Draco Nebula. In both cases, we have found a good correlation between the signal in the FUV band (1350 -- 1750 \AA) and the NUV band (1750 -- 2850 \AA) but with an additional component in the FUV which is not seen in the NUV. This was identified as fluorescent emission from the Lyman band of molecular hydrogen in Region I and in the nearby cloud LVC 88+36-2 in these observations, where the ratio was correlated with the H~{\small I} column density. However, there was excess emission throughout the Draco region which was not correlated (or anti-correlated) with N(H~{\small I}) and this may be due either to H$_{2}$ emission or to line emission from hot gas. While \galex\ observations are invaluable in probing the diffuse background at unprecedented sensitivity and spatial resolution, spectra will still be necessary to fully understand the observations. However, we strongly recommend that the FUV/NUV ratio can be used to identify the atomic and molecular emission regions in the \galex\ survey fields all over the sky. 

The scattered light from the interstellar dust is consistent with an optically thin layer in the Draco region transitioning to optically thick in the earlier Region I results, although the thermal emission in the infrared is optically thin in both cases. The $F_{UV}/F_{IR}$ ratio follows an exponential curve across both regions, as would be expected for optically thick media. Interestingly, the $F_{UV}/F_{IR}$ ratio in Orion follows exactly the same curve even though both the UV and IR values are higher by a factor of almost 200 due to the intense radiation field. In general, we find that the $F_{UV}/F_{IR}$ ratio strongly depends on the local effects such as the proximity of hot stars to the scattering medium and its  optical depth.

We have determined optical constants $a$(0.45~$\pm$~0.08) and $g$(0.56~$\pm$~0.10) in the NUV band and $a$(0.32~$\pm$~0.09) and $g$(0.51~$\pm$~0.19) in the FUV band for the dust in Draco, largely consistent with previous observational and theoretical determinations \citep{Gord04}. Regardless of the actual value of the optical constants, we find that the ratio between the FUV and the NUV dust scattered light is 0.8 over a wide range of optical depths (Draco and Region I). We have also estimated the extragalactic contribution of 58~$\pm$~18 \phunit\ in the NUV band using our model, which is in good agreement with the derived limit of 49~$\pm$~13~\phunit\ for the band using the {\it Spitzer}~FLS sources. This gives strong evidence that most of the diffuse background derived from the \galex\ observations have a Galactic origin specifically at galactic latitudes, $\left |~b~\right |~<~40\degr$.

We have begun a massive program to look at the small scale structure of diffuse background in all \galex\ data of greater than 5000 seconds. These include data throughout the sky and sample a variety of different environments, although avoiding bright UV regions such as the Coalsack or Orion. In parallel, we are developing more sophisticated models to better match the high quality data obtained here. We believe the \galex\ data will allow us to place the study of the diffuse UV radiation on the same level as \iras\ did for the infrared cirrus.

\acknowledgements
We thank our anonymous referee for the helpful comments and suggestions which have significantly improved this paper. This research is based on data from the NASA's \galex\ program. \galex\ is operated for NASA by the California Institute of Technology under NASA contract NAS5-98034. We have also made use of NASA's Astrophysics Data System and the SIMBAD database operated at CDS, Strasbourg, France. We acknowledge the use of NASA's SkyView facility (http://skyview.gsfc.nasa.gov) located at NASA Goddard Space Flight Center. 

NVS is supported by a DST Young Scientist award. Support for RCH was provided by NASA \galex\ grant NNGO5GF19G to the Johns Hopkins University. 

As always, Patrick Morrissey has greatly helped with our understanding of the GALEX data.

{\it Facilities:} \facility{\galex}.

\clearpage

\begin{deluxetable}{l c c c c c c c c c c}
\rotate
\tabletypesize{\scriptsize}
\tablenum{1}
\tablecaption{Observation Log}
\tablewidth{0pt}
\tablehead
{\colhead{Tile Name} & \colhead{RA} & \colhead{Dec} & \colhead{$l$} & \colhead{$b$} & \colhead{NUV Exposure} & \colhead{FUV Exposure\tablenotemark{a}} & \colhead{Observation period} & \colhead{NUV Visits} & \colhead{FUV Visits\tablenotemark{a}}\\
\colhead{} & \colhead{(deg)} & \colhead{(deg)} & \colhead{(deg)} &\colhead{(deg)} & \colhead{(sec)} &\colhead{(sec)} & \colhead{(yyyy/mm/dd)} & \colhead{} &\colhead{}
}
\startdata
SIRTFFL-00 & 259.11 & 59.91	& 88.84	& 35.05	& 52917.15 & 52016.95 & 2003/07/03 -- 2008/08/25 & 41 & 39 \\
SIRTFFL-01 & 260.41 & 59.34	& 88.08	& 34.44	& 26006.10 & 30922.1 & 2003/07/04 -- 2004/07/26 & 20 & 24 \\
SIRTFFL-02 & 260.09 & 58.5	& 87.08	& 34.66	& 39037.05 & 26859.35 & 2003/08/18 -- 2007/09/02 & 30 & 25 \\
SIRTFFL-03 & 258.33 & 58.86	& 87.61	& 35.55	& 39830.40 & 29570.9 & 2003/08/19 -- 2007/09/01 & 30 & 28 \\
SIRTFFL-04 & 256.98 & 59.72	& 88.76	& 36.13	& 3874.45 & 3874.45 & 2004/05/01 -- 2004/05/01 & 3 & 3	\\
SIRTFFL-05 & 260.68 & 60.7	& 89.71	& 34.2	& 5305	 & 5305	  & 2004/05/01 -- 2004/05/03 & 4 & 4	\\
SIRTFFL-06 & 257.58 & 60.45	& 89.6	& 35.74	& 27658.75 & 2540.55  & 2005/07/25 -- 2008/04/07  & 22 & 2	\\
SIRTFFL-07 & 260.54 & 60.81	& 89.85	& 34.26	& 34376.55& 21276.1 & 2005/07/27 -- 2007/09/02  & 25 & 20	\\
SIRTFFL-08 & 262.61 & 59.15	& 87.78	& 33.33	& 40639.6 & 22540  & 2005/06/19 -- 2007/09/01  & 28 & 20	\\
SIRTFFL-09 & 257.2 & 59.72	& 88.74	& 36.02	& 15737.4 & 2755.45  & 2005/07/29 -- 2008/04/07  & 12 & 2	\\
SIRTFFL-10 & 256.99 & 58.8 &	87.63	& 36.24	& 27383.75& 10757.7 & 2005/07/25 -- 2007/08/28  & 21 & 15	\\
Region I\tablenotemark{b} & 142.04 & 70.39	& 142.3	& 38.2	& 35210	 & 14821  & 2005/01/11 -- 2007/01/04  & 22 & 10	\\
%\hline
\enddata
\label{obs_log}
\tablenotetext{a}{There are often fewer visits in the FUV because of intermittent failures in the FUV power supply.}
\tablenotetext{b}{Tile name: GI1-005007-J092810p702308.}
\end{deluxetable}
\clearpage

\begin{deluxetable}{l c c c c r}
\tablenum{2}
\tablecaption{H~{\small I} Components in the Field}
\tablewidth{0pt}
\tablehead
{\colhead{Cloud} & \colhead{$l$} & \colhead{$b$} & \colhead{Peak N(H~{\small I})} & V$_{LSR}$\\
\colhead{} & \colhead{(deg)} & \colhead{(deg)} & \colhead{(cm$^{-2}$)} & (km s$^{-1}$)
}
\startdata
Ridge (LVC) &  87.44 & 35.93 & 1.8 $\times$ 10$^{20}$ & -2\\
IVC1        &89.47 & 34.25 & 5.6 $\times$ 10$^{19}$ & -41\\
IVC2        &88.82 & 34.17 & 5.0 $\times$ 10$^{19}$ & -41\\
IVC3        &86.53 & 33.73 & 4.4 $\times$ 10$^{19}$ & -34\\
IVC4 (Draco) & 89.85 & 35.60 & 7.8 $\times$ 10$^{19}$ & -23\\
HVC (Complex C) &89.15 & 35.20 & 6.9 $\times$ 10$^{19}$ & -190
\enddata
\label{h1_comp}
\end{deluxetable}

\clearpage

\begin{deluxetable}{l c c c c c c}
%%\tabletypesize{(11pt)}
\tablenum{3}
\tablecaption{Airglow and Zodiacal Contribution in Each Field}
\tablewidth{0pt}
\tablehead
{\colhead{Tile Name} & \multicolumn{2}{c}{Average Airglow} & \colhead{Zodiacal light} & \multicolumn{2}{c}{Total Foreground Emission\tablenotemark{a}}\\
\colhead{} & \colhead{FUV} & \colhead{NUV} & \colhead{NUV} & \colhead{FUV} & \colhead{NUV}\\
\colhead{} & \multicolumn{5}{c}{(\phunit)}
}
\startdata
SIRTFFL-00 & 391 & 394 & 367 & 396  & 766 \\
SIRTFFL-01 & 387 & 338 & 407 & 392  & 750 \\
SIRTFFL-02 & 332 & 314 & 373 & 337  & 692 \\
SIRTFFL-03 & 318 & 308 & 382 & 323  & 695 \\
SIRTFFL-04 & 320 & 378 & 342 & 325  & 725 \\
SIRTFFL-05 & 362 & 440 & 342 & 367  & 787 \\
SIRTFFL-06 & 306 & 304 & 358 & 311  & 667 \\
SIRTFFL-07 & 356 & 313 & 381 & 361  & 699 \\
SIRTFFL-08 & 327 & 304 & 365 & 332  & 674 \\
SIRTFFL-09 & 333 & 333 & 367 & 338  & 705 \\
SIRTFFL-10 & 368 & 369 & 365 & 373  & 739 \\
Region I\tablenotemark{b} & 349 & 355 & 440 & 354  & 800 
\enddata
\label{ag_zl}
\tablenotetext{a}{Includes 5 \phunit\ dark count.}
\tablenotetext{b}{Tile name: GI1-005007-J092810p702308.}
\end{deluxetable}

\clearpage

\begin{deluxetable}{l c c c c c c}
%%\tabletypesize{(11pt)}
\tablenum{4}
\tablecaption{Correlation details of UV emission with different components of N(H~{\small I})}
\tablewidth{0pt}
\tablehead
{\colhead{Data} & \colhead{LVC} & \colhead{IVC}&\colhead{HVC} & \colhead{LVC+IVC}&\colhead{LVC+IVC+HVC}\\
}
\startdata
FUV Ridge & 0.88 & 0 & 0 & 0.84 & 0.82 \\
FUV Total &  0.75 & 0.3 & -0.09 & 0.70 & 0.51  \\
NUV Ridge & 0.63 & 0 & 0 & 0.63 & 0.63  \\
NUV Total &  0.63 & 0.27 & -0.05 & 0.63 & 0.52  \\
\enddata
\label{correl}
\end{deluxetable}
\clearpage

\begin{deluxetable}{l c c c c c c}
%%\tabletypesize{(11pt)}
\tablenum{5}
\tablecaption{Number Counts of Extragalactic Objects present in the NUV diffuse map of {\it Spitzer} field}
\tablewidth{0pt}
\tablehead
{\colhead{AB Mag} & \colhead{N$_{objects}$} & \colhead{Log (N$_{objects}$/deg$^{2}$/mag)}\\
}
\startdata
20.25 & 14 & 0.94 \\
20.75 & 130 & 1.91  \\
21.25 & 584 & 2.56  \\
21.75 & 1941 & 3.08  \\
22.25 & 3288 & 3.31 \\
22.75 &  4211 & 3.42  \\
23.25 & 4307 & 3.43  \\
23.75 &  3904 & 3.39  \\
24.25 & 3364 & 3.32  \\
\enddata
\label{num_cts_tab}
\end{deluxetable}

\clearpage
\begin{figure}
\epsscale{0.7}
\plotone{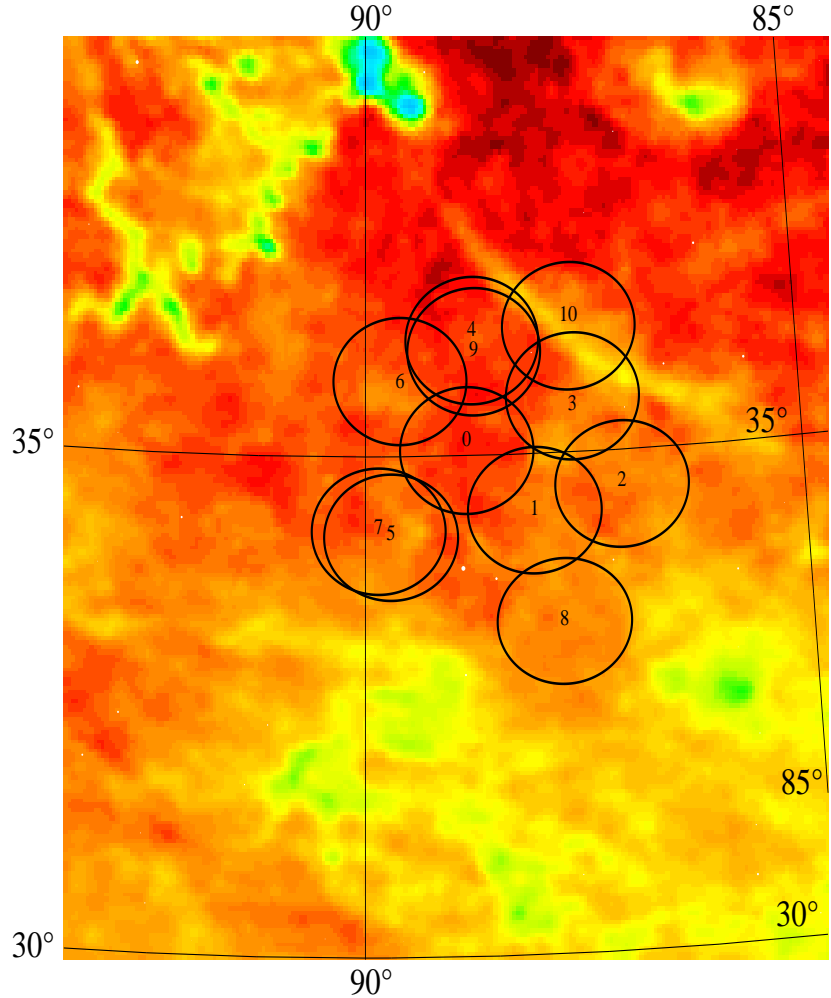}
\figcaption{\iras\ map of the region in galactic coordinates. The GALEX field of view of the 11 DIS targets are overplotted as circles with diameter 1.25\degr\ and marked as 0 to 10. The bright arc  extending through the fields 3 \& 10 is the low velocity cloud (LVC 88+36-2) discussed in the text and the brightest feature on the left top is the Draco Nebula.\label{IR_img}}
\end{figure}

\clearpage
\begin{figure}
\epsscale{0.7}
\plotone{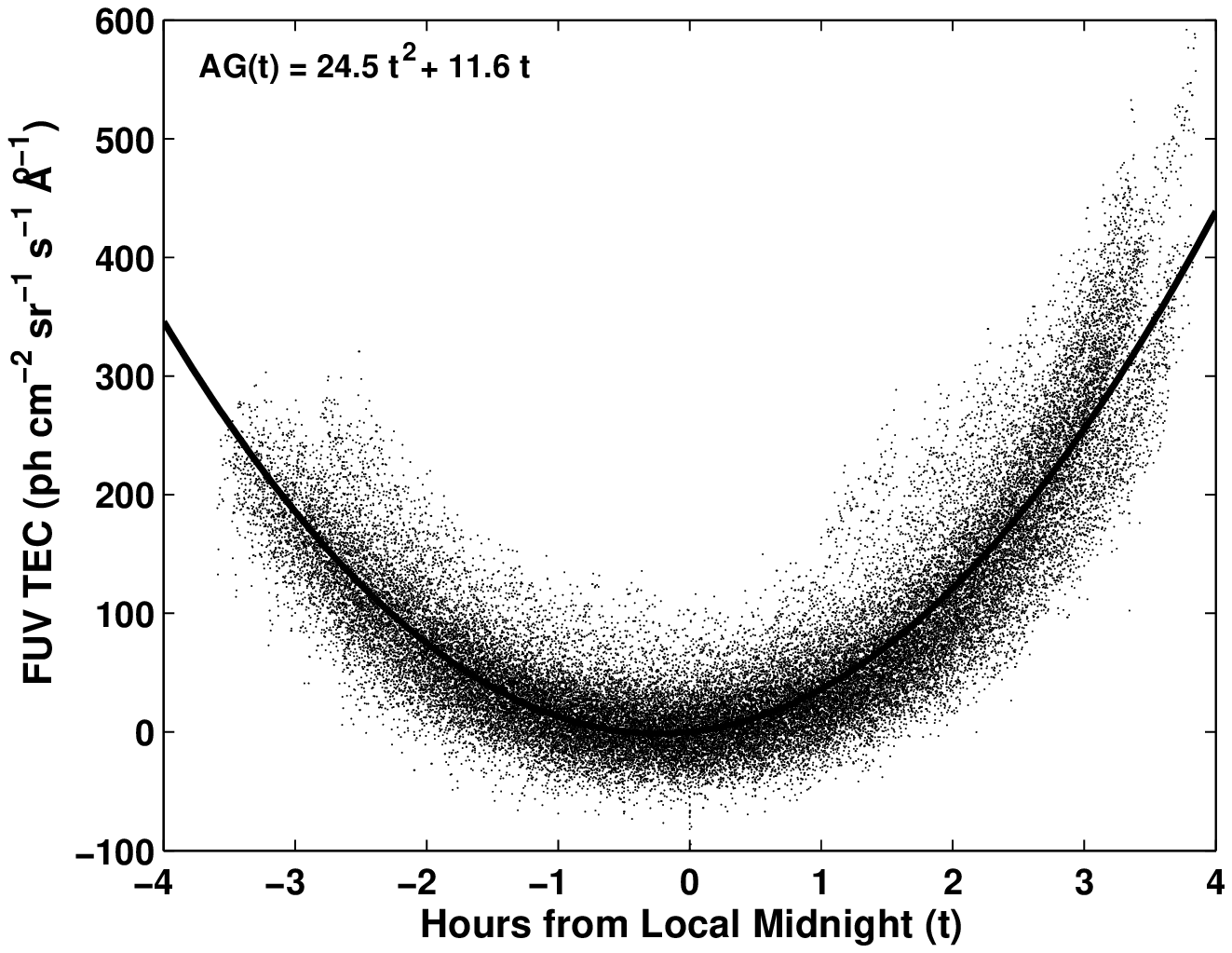}
\plotone{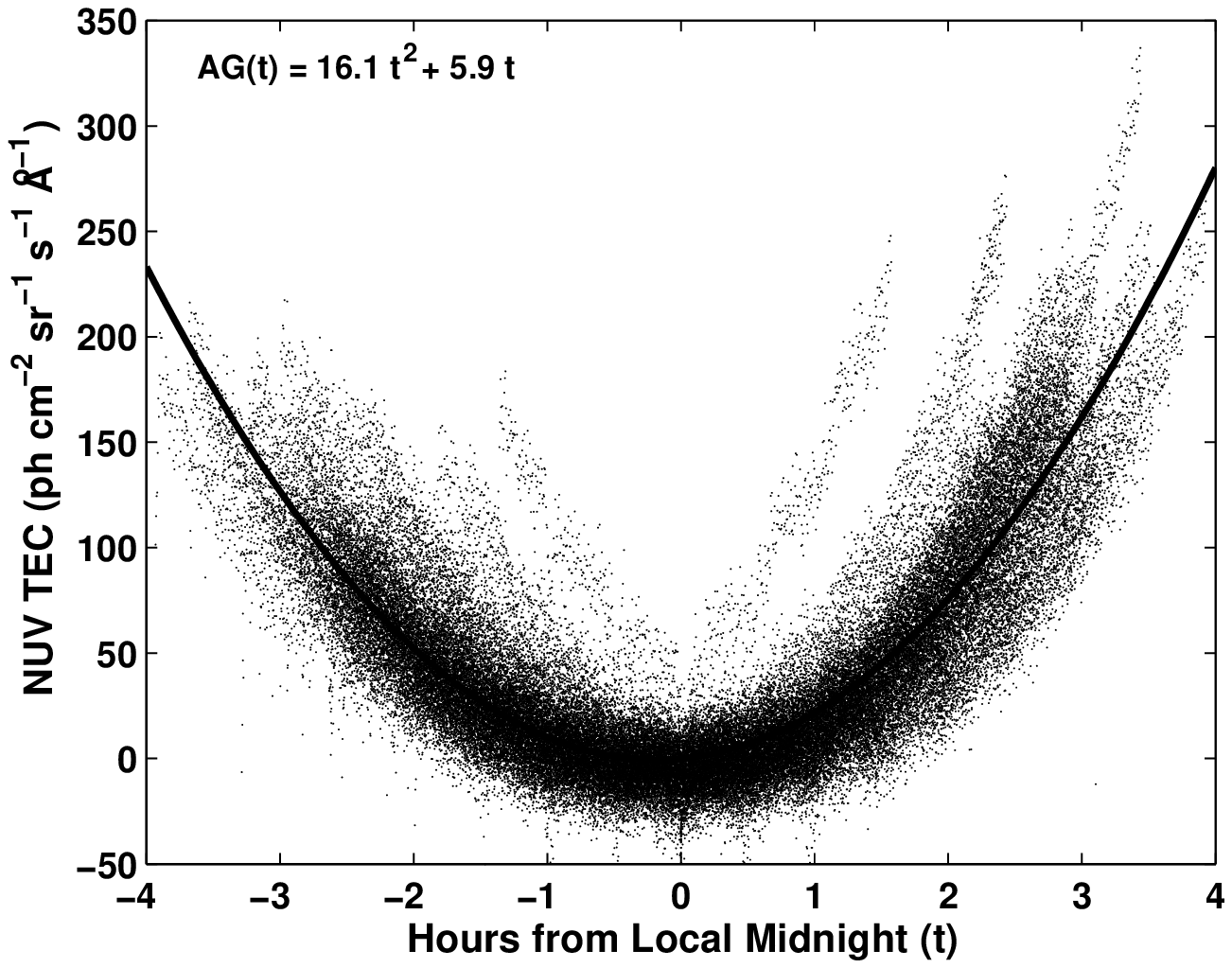}
\figcaption{Total count rate (TEC, in \phunit) in the FUV (top) and NUV (bottom) is plotted against the local time from midnight. A baseline has been subtracted from each visit so that the count rate is zero at local midnight. The solid line represents the best fit curve to the data whose quadratic equation is given in the left top of the plot. \label{TEC_time}}
\end{figure}

\clearpage
\begin{figure}
\epsscale{0.8}
\plotone{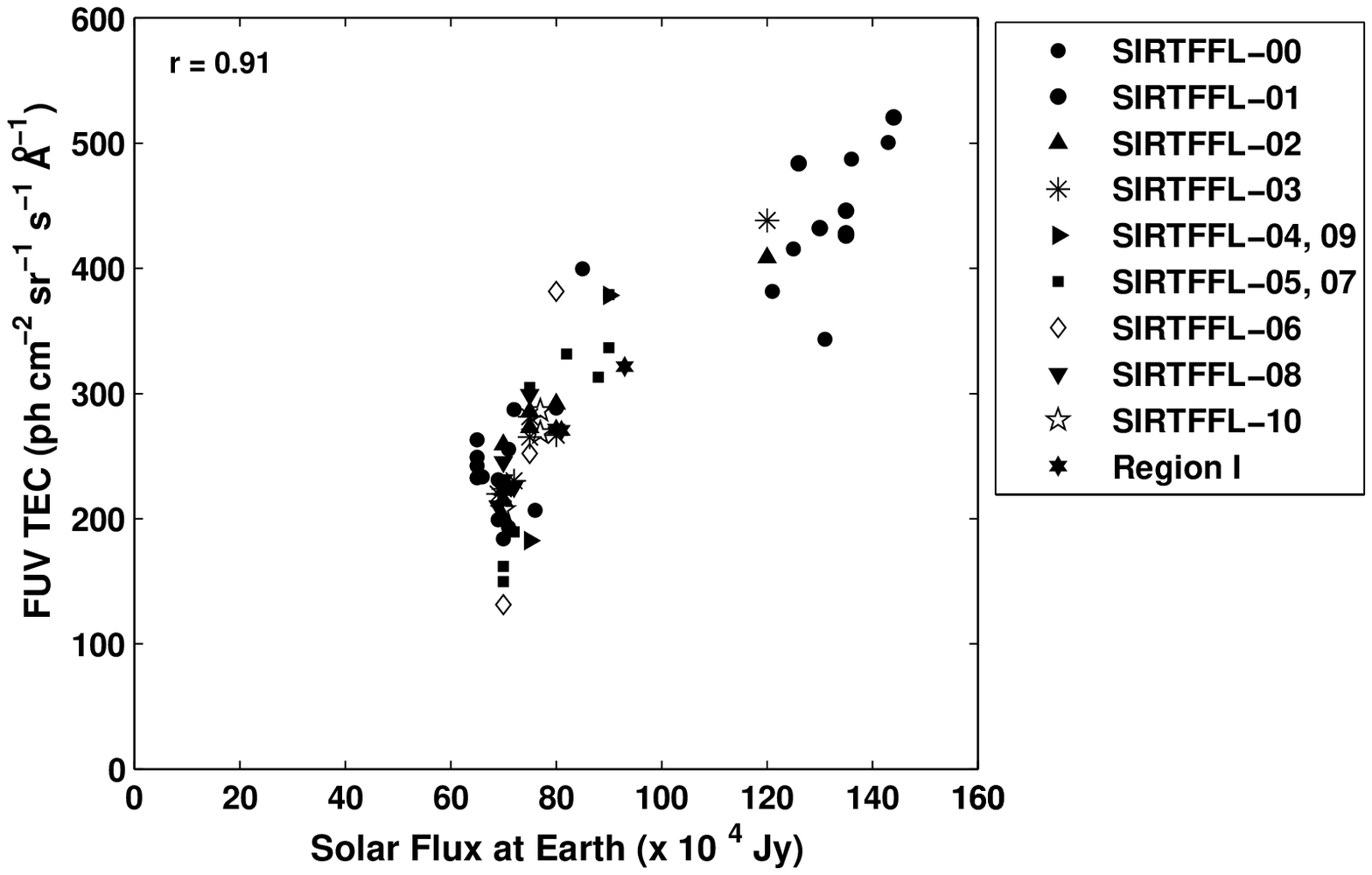}
\plotone{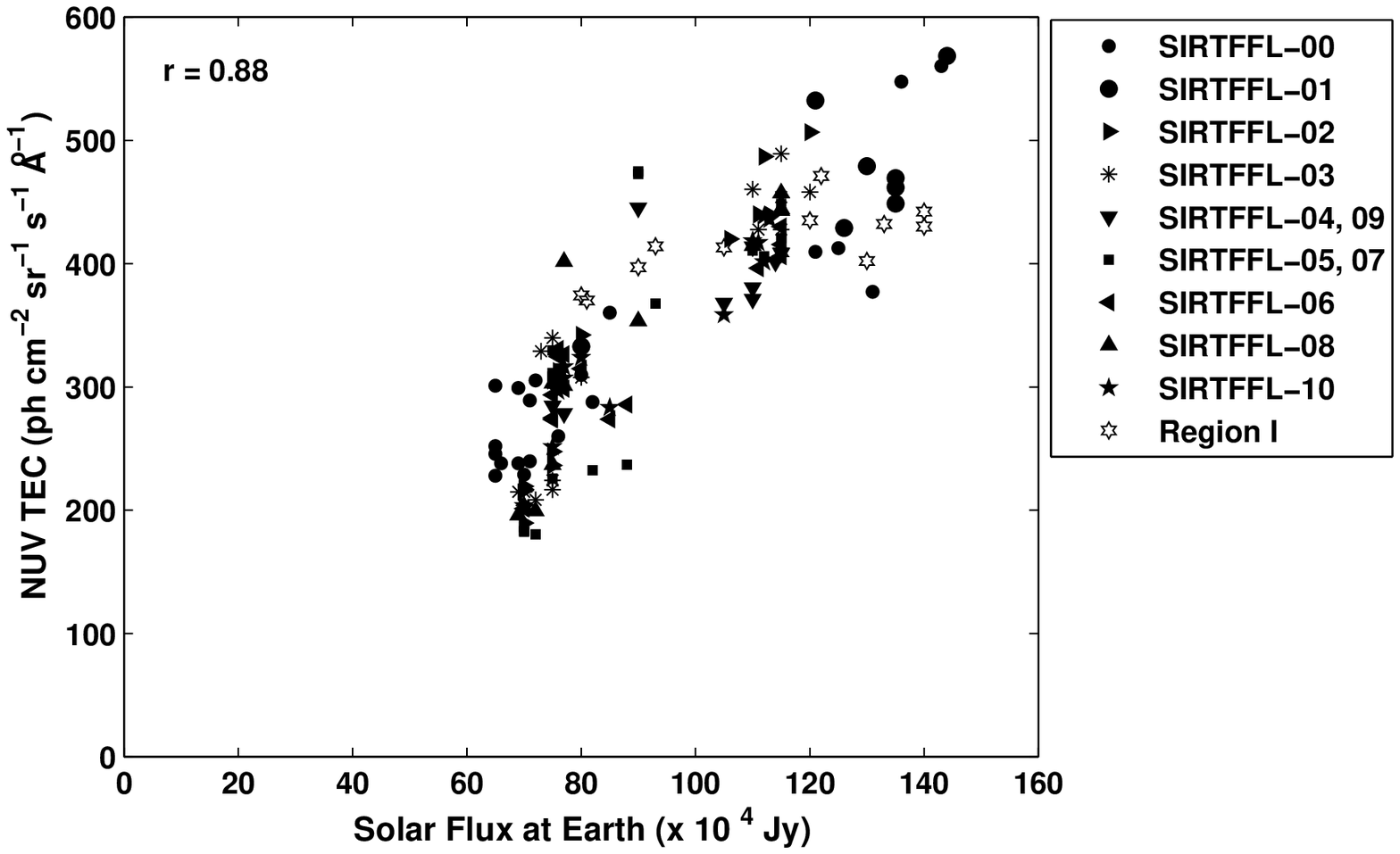}
\figcaption{Minimum TEC level in each visit at the local midnight is plotted against the 10.7 cm solar flux at the Earth for FUV (top) and NUV (bottom) channel. An offset was subtracted from each observation (one set of visits). The strong correlation observed here indicates that the variation in the TEC level within an observation is due to the solar activity.\label{solar_flux}}
\end{figure}

\clearpage
\begin{figure}
\epsscale{0.8}
\plotone{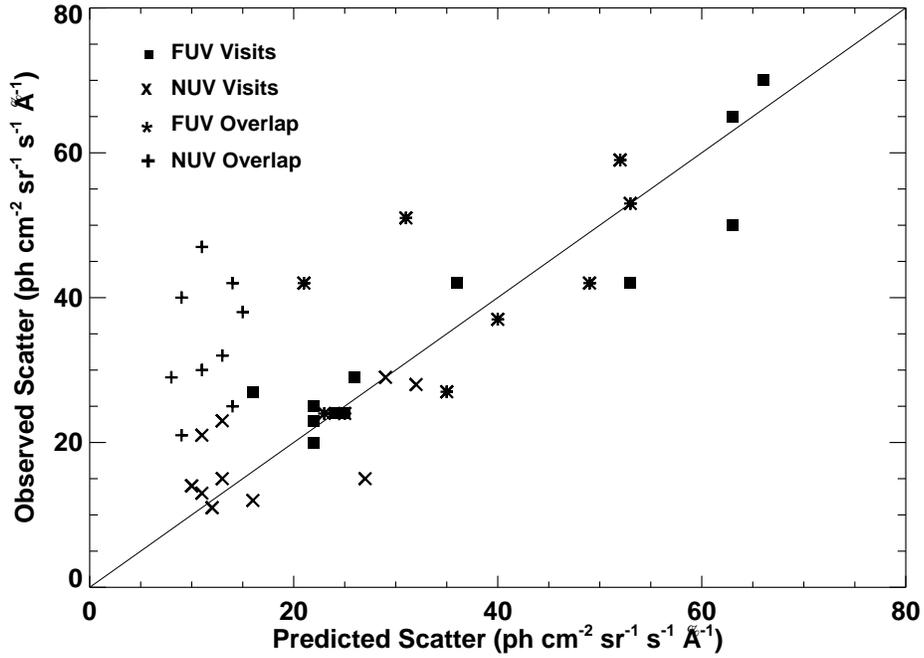}
\figcaption{Comparison of observed scatter and intrinsic photon noise in the data. The symbols `filled square' and `x' represent the scatter when a single observation is broken up into two sets of visits for each band, respectively, while the symbols `asterisk' and `+' represent the scatter in the regions of overlap between different observations. In general, the observed scatter is consistent with photon noise alone with the high points being due to the smaller area of overlap.\label{scat_plot}}
\end{figure}

\clearpage
\begin{figure}
\epsscale{0.9}
\plotone{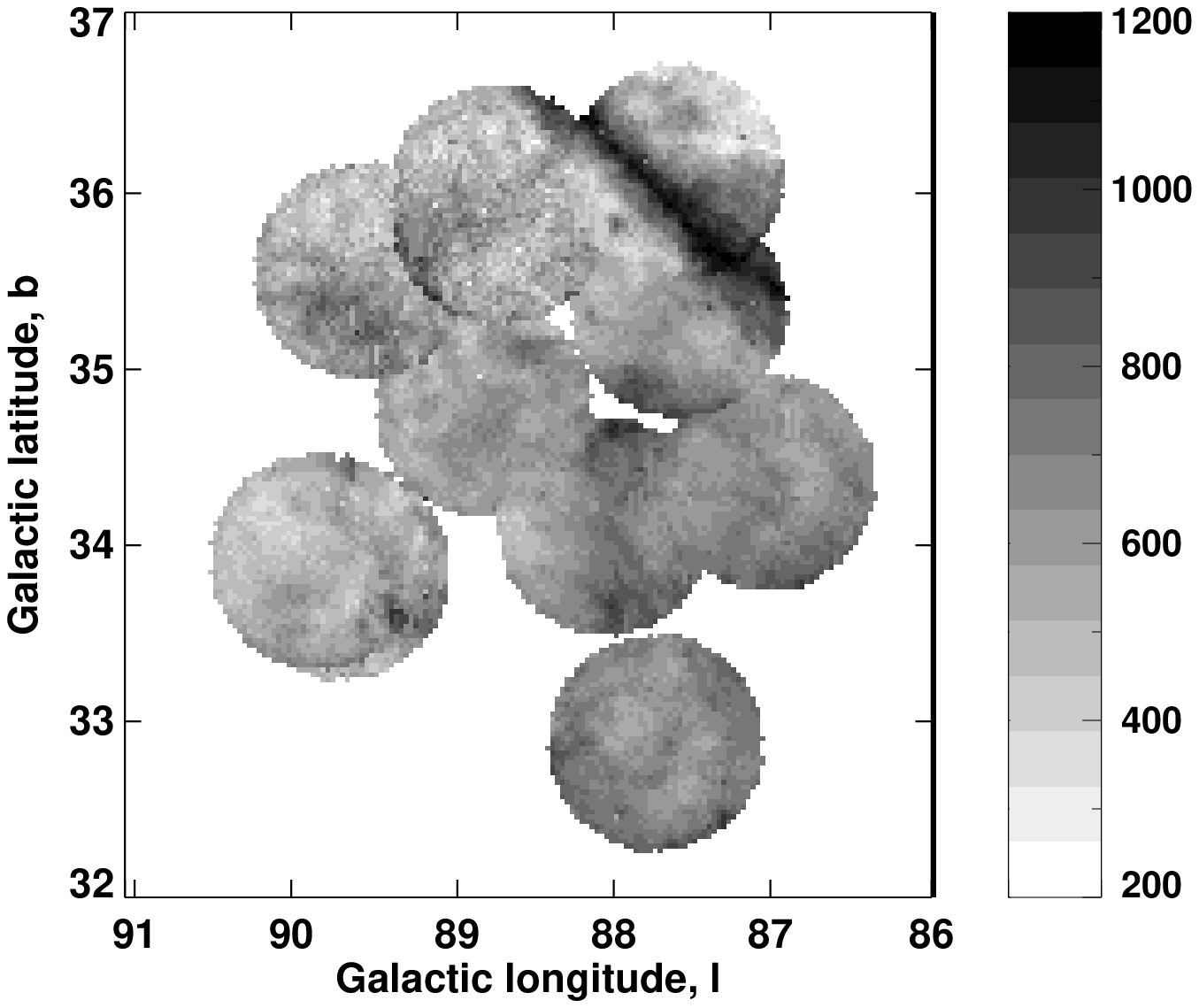}
\plotone{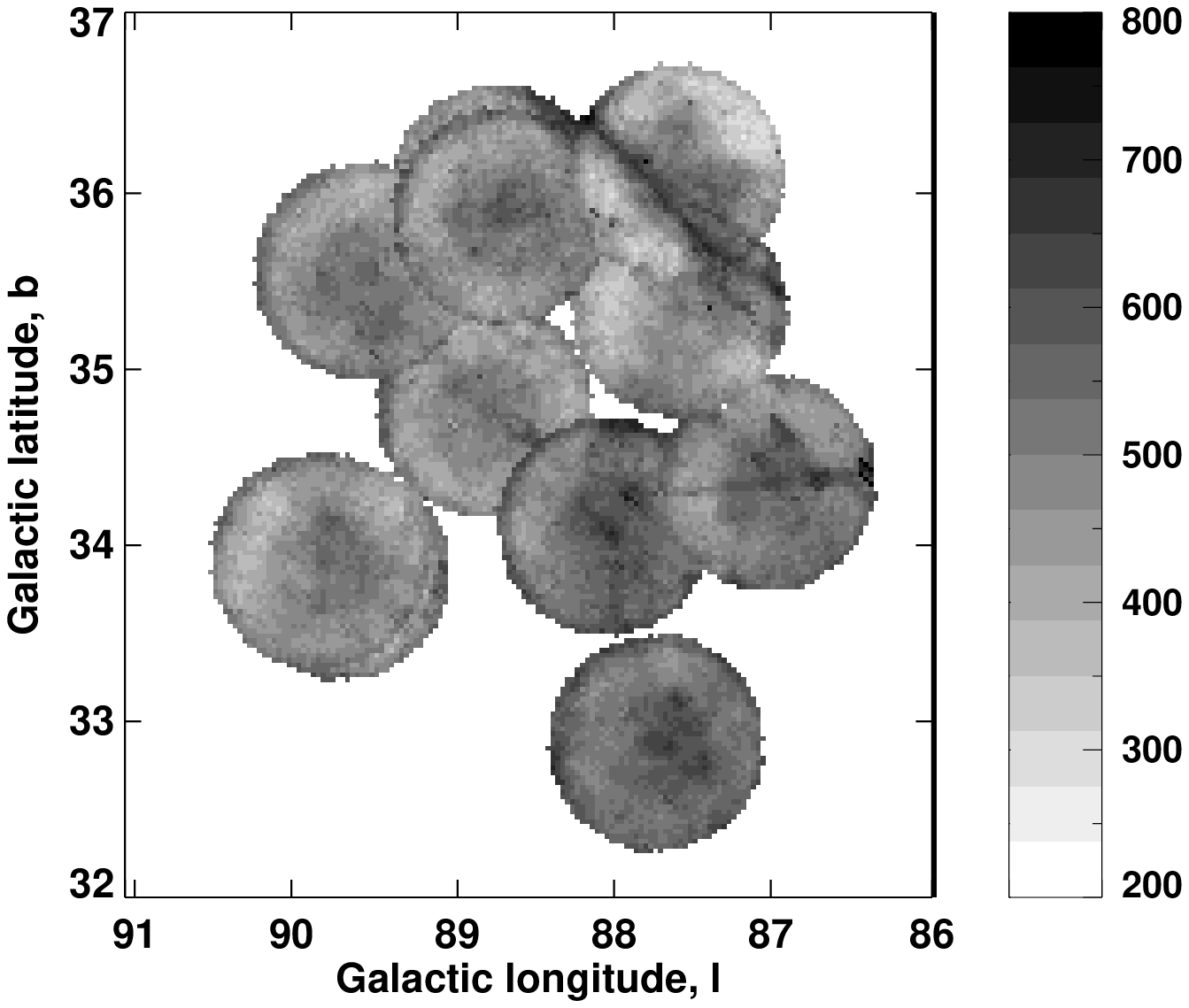}
\figcaption{Diffuse FUV (top) \& NUV (bottom) images (in \phunit) of the region derived from the central 1.2\degr\ field of view of each \galex\ observation. The foreground emission has already been subtracted from each band. \label{diffuse_image}}
\end{figure}

\clearpage
\begin{figure}
\epsscale{0.8}
\plotone{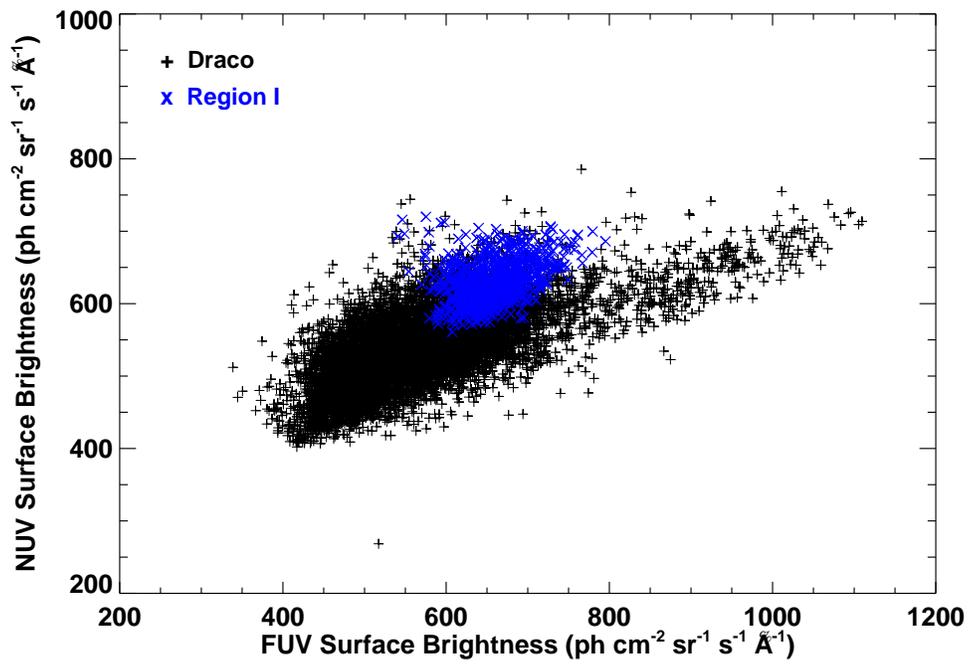}
\figcaption{Correlation between FUV and NUV intensity. The blue points (`x') represent Region~I and the black points `+' represent the Draco region. Good correlation between the FUV and NUV bands indicating that the dominant contributor of the diffuse background in the field is the scattered starlight from the interstellar dust grains.\label{fuv_nuv}}
\end{figure}

\clearpage
\begin{figure}
\epsscale{0.7}
\plotone{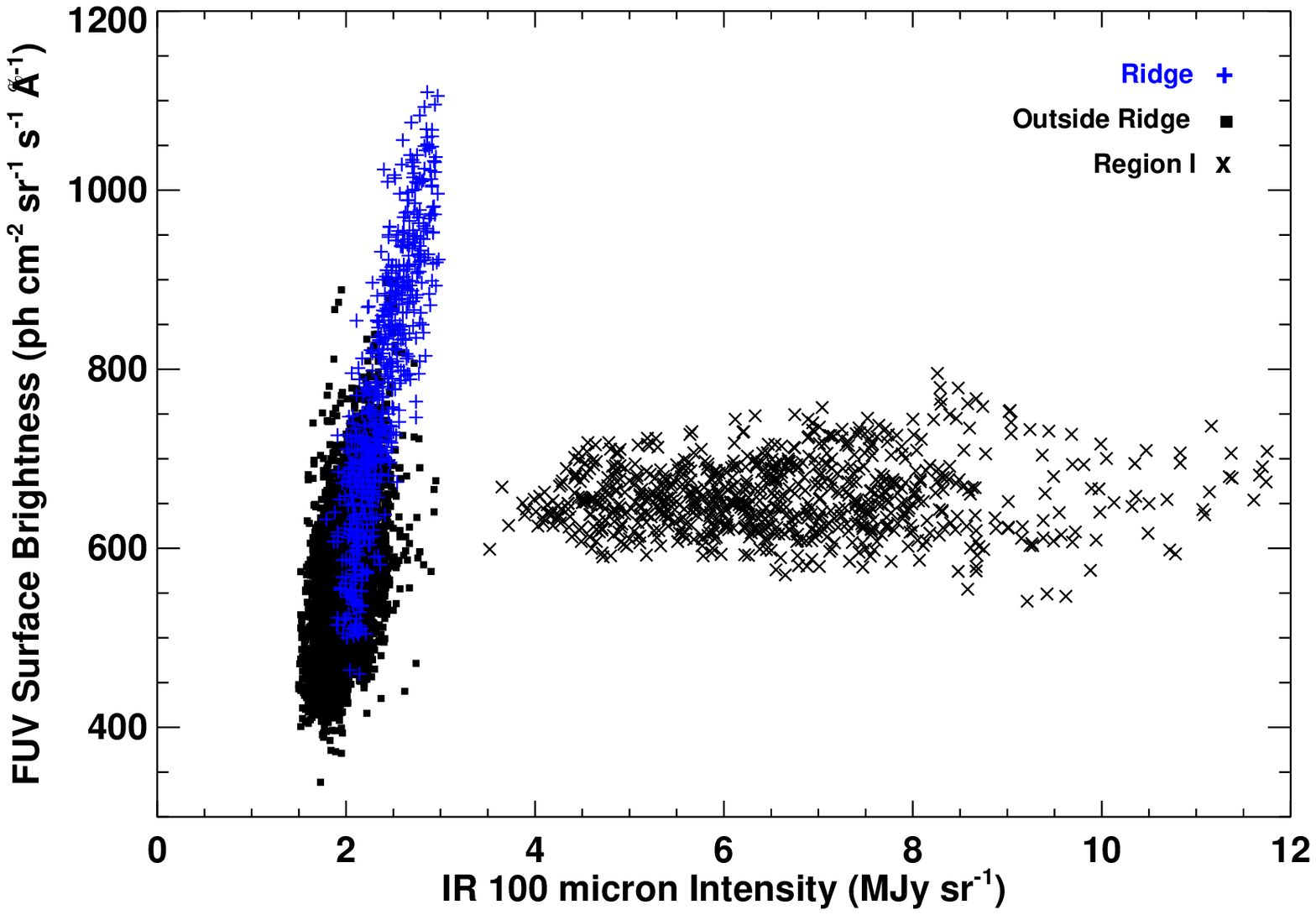}
\plotone{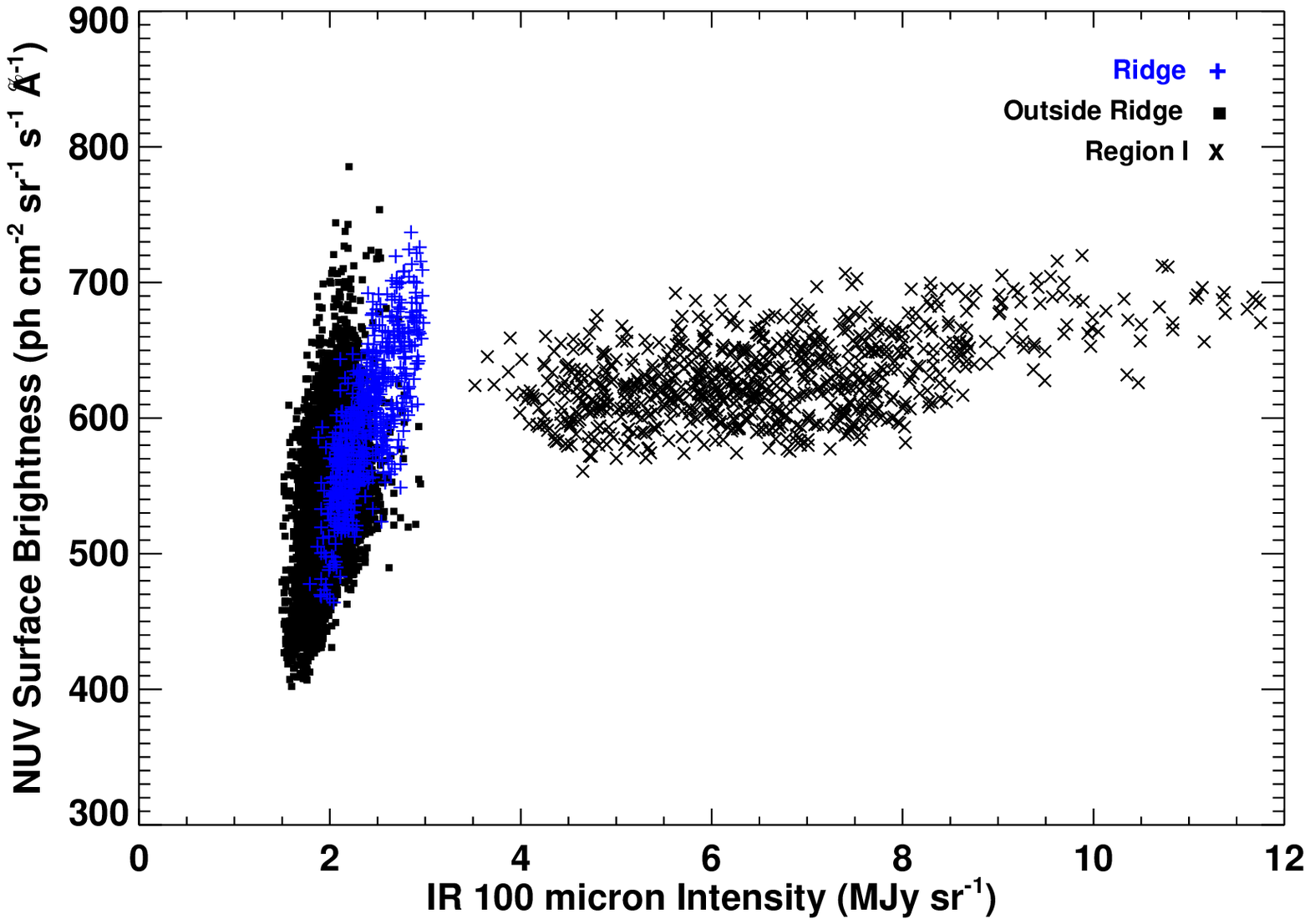}
\figcaption{Correlation between \iras\ 100 micron intensity and diffuse FUV (top) and NUV (bottom) background radiation. In each plot, the blue points (`+') represent the ridge of dust, the `dots' represent the region outside the ridge and the `x' points represent Region~I. The background radiation is strongly correlated with IR in Draco region but is saturated in Region~I because of the high optical depth in the UV.\label{UV_IR}}
\end{figure}

\clearpage
\begin{figure}
\epsscale{0.8}
\plotone{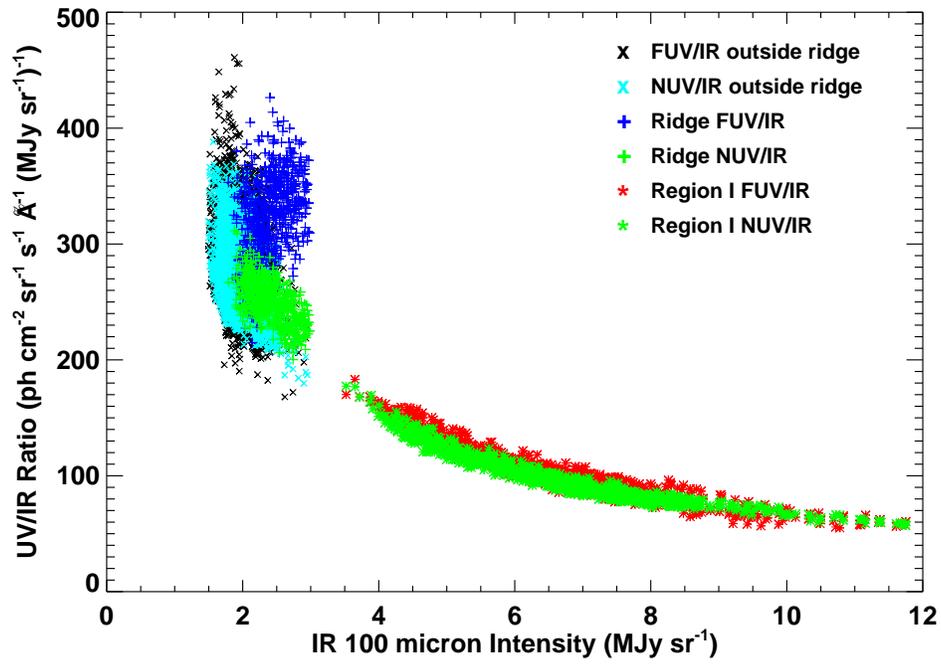}
\figcaption{UV/IR ratio (in \phunit(MJy sr$^{-1}$)$^{-1}$) as a function of IR 100 \micron\ intensity. The ratio exponentially drops off with IR due to the rapid increase of optical depth in UV.\label{UV_IR_ratio}}
\end{figure}

\clearpage
\begin{figure}
\epsscale{0.8}
\plotone{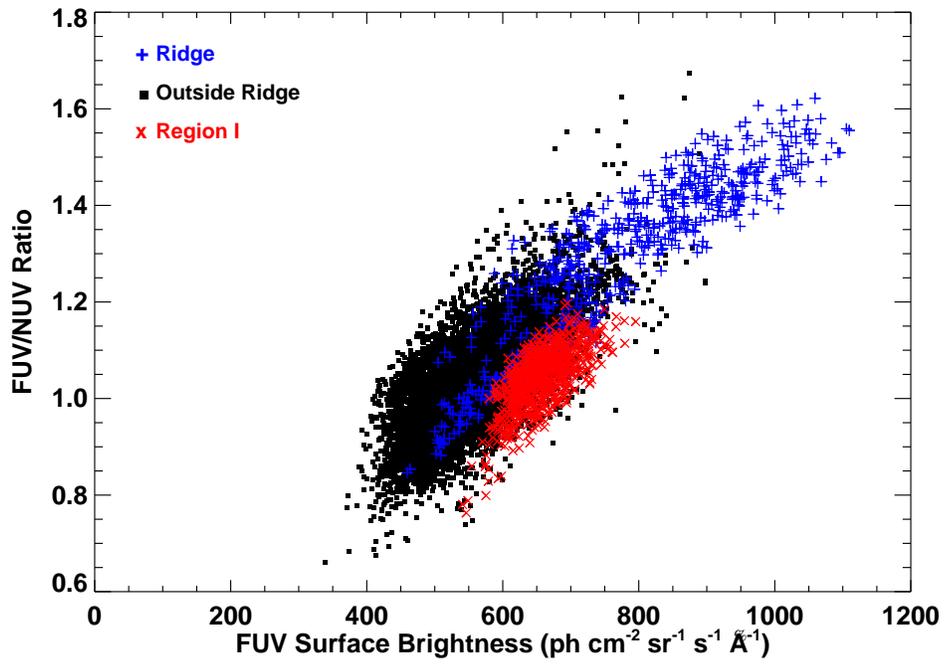}
\figcaption{Ratio between the UV bands (after subtracting the foreground emissions) is plotted against the FUV surface brightness. The increase in the ratio with FUV radiation indicates the presence of excess emission in the FUV band. \label{ratio_fuv}}
\end{figure}

\clearpage
\begin{figure}
\epsscale{0.9}
\plotone{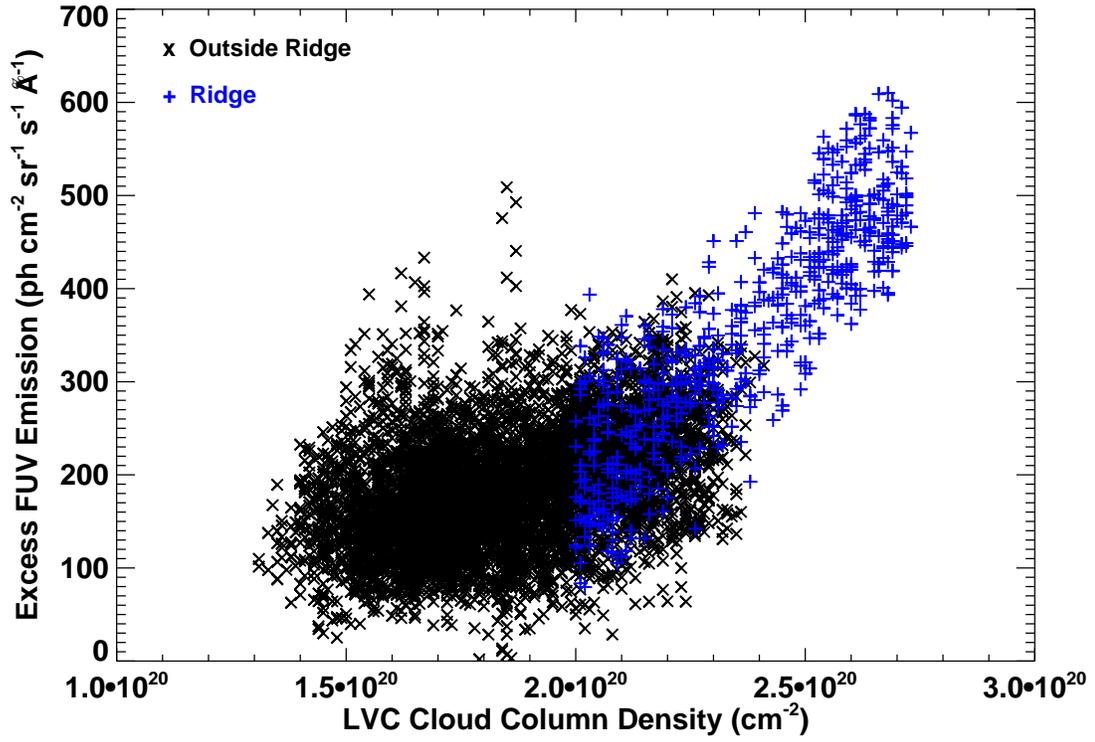}
\figcaption{Excess FUV emission in the observations is plotted against N(H~{\small I}) in the LVC \citep{Lockman05}. There is a strong correlation inside the dust ridge where the excess emission is due to molecular hydrogen fluorescence but a poorer correlation outside where the excess emission may be due to line emission from C~{\small IV} or Si~{\small II}. \label{lvc-h2f}}
\end{figure}

\clearpage
\begin{figure}
\epsscale{0.9}
\plotone{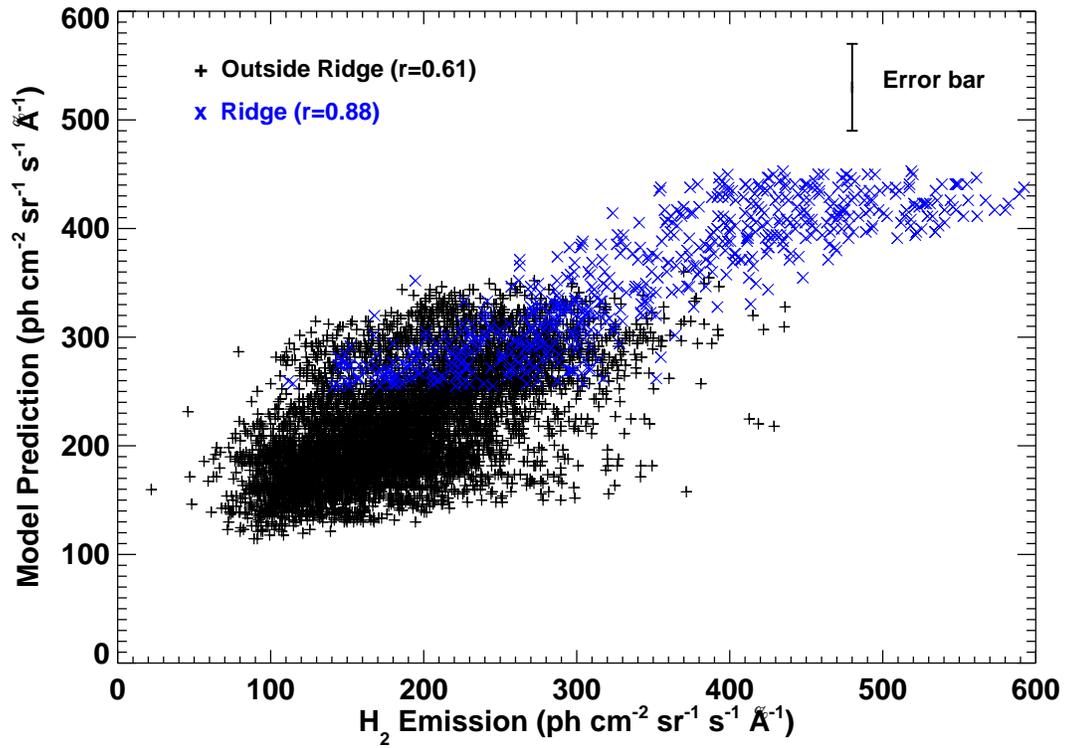}
\figcaption{Predicted levels of  H$_{2}$ emission with a formation rate (R) of 1 x 10$^{-17}$ cm$^{-3}$ s$^{-1}$ are plotted against the excess emission in the field. There is reasonable agreement everywhere but particularly in the nearby cloud LVC 88+36-2.\label{H2_ridge}}
\end{figure}

\clearpage
\begin{figure}
\epsscale{0.8}
\plotone{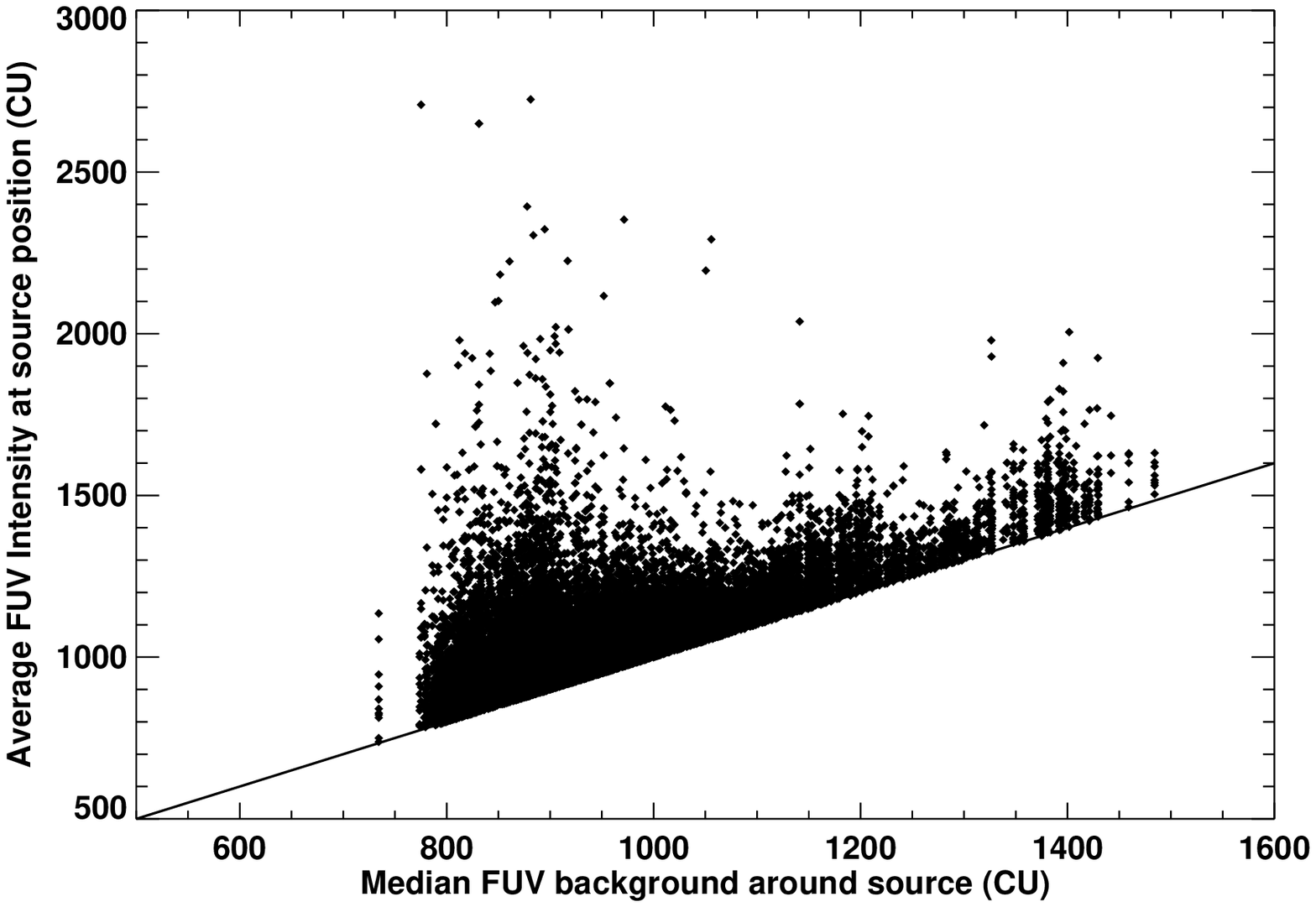}
\plotone{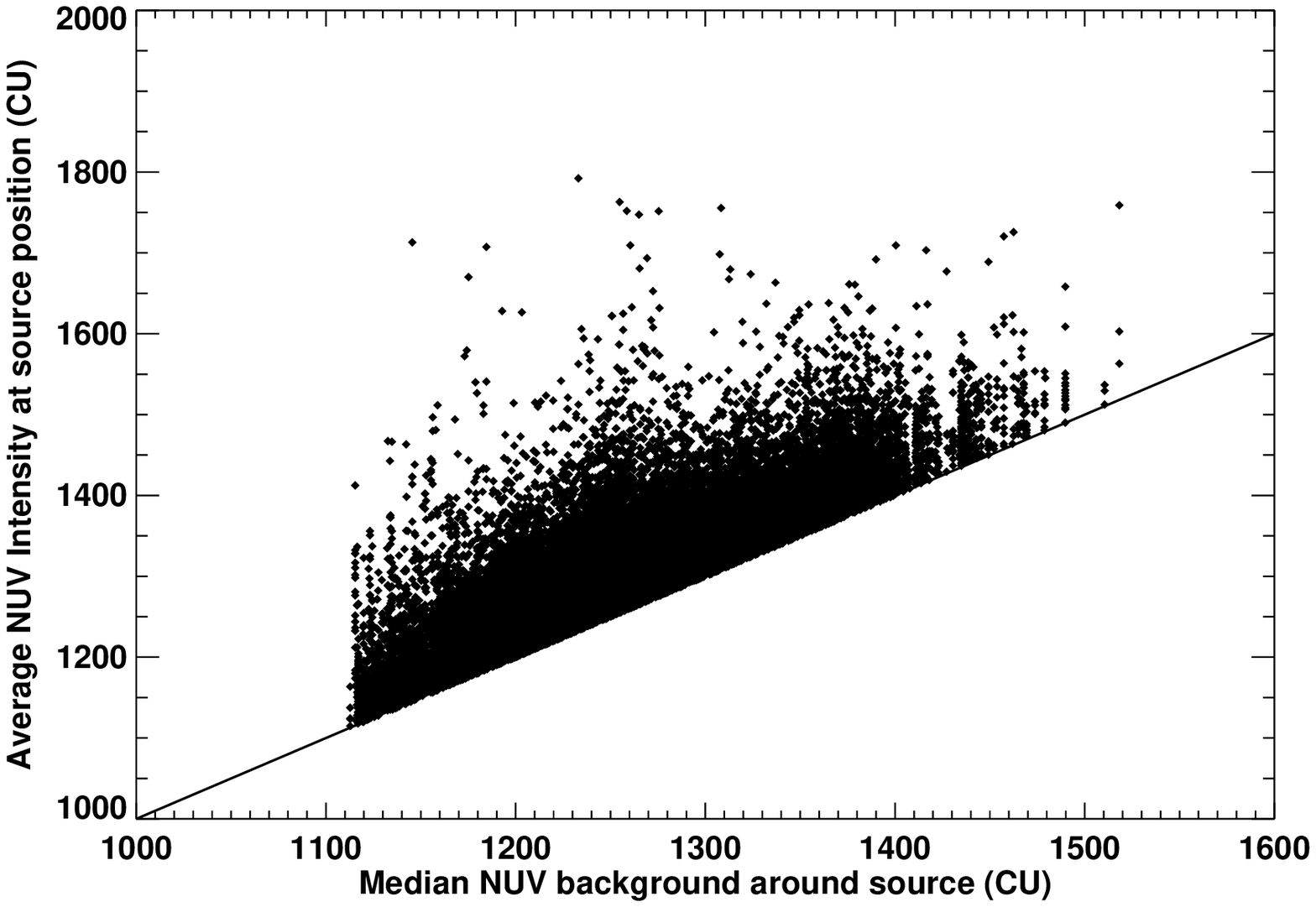}
\figcaption{Comparision of average UV intensity (in continum unit; 1 CU = 1\phunit) 
from 9\arcsec\ bin and median background from 2\arcmin\ bin centered at each {\it IRAC} 
object position in our diffuse maps. The enhancement in some of the {\it IRAC} source position 
indicate the presence of undetected faint galaxies by the SExtractor in our diffuse maps.\label{uv_enh}}
\end{figure}
\clearpage
\begin{figure}
\epsscale{0.8}
\plotone{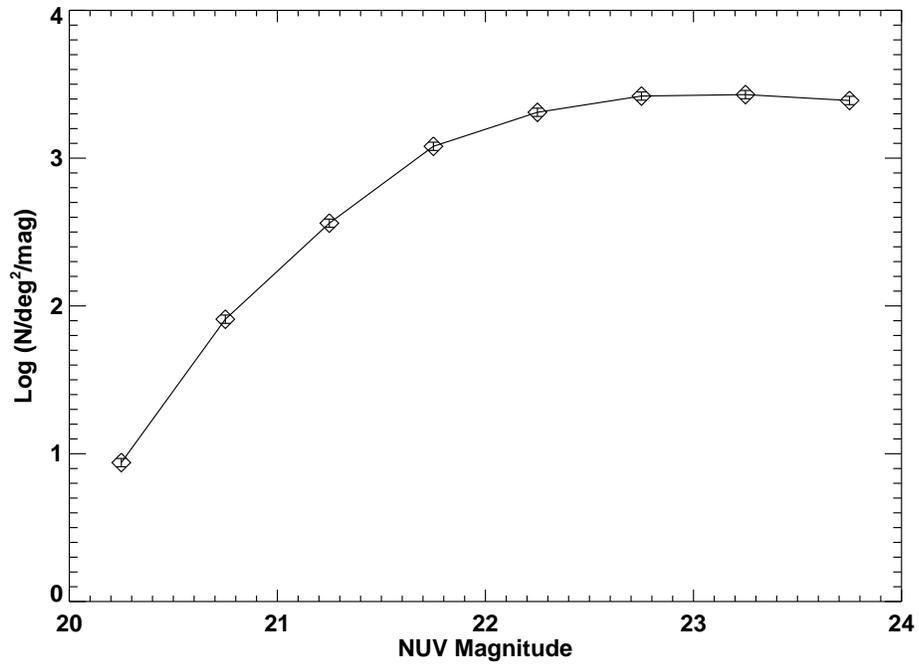}
\figcaption{Number counts of extragalactic objects present in the diffuse NUV map of the {\it Spitzer} field. The solid line is the best-fit curve.\label{num_cts}}
\end{figure}

\end{document}